# Ferromagnetism of a graphite nodule from the Canyon Diablo meteorite

J. M. D. Coey*, M. Venkatesan*, C. B. Fitzgerald*, A. P. Douvalis* and I. S. Sanders[†]

*Physics Department, Trinity College, Dublin 2, Ireland.*

[†]*Geology Department, Trinity College, Dublin 2, Ireland.*

**There have recently been various reports of weak ferromagnetism in graphite[1,2] and synthetic carbon materials[3] such as rhombohedral $C_{60}$[4], as well as a theoretical prediction of a ferromagnetic instability in graphene sheets[5]. With very small ferromagnetic signals, it is difficult to be certain that the origin is intrinsic, rather than due to minute concentrations of iron-rich impurities. Here we take a different experimental approach to study ferromagnetism in graphitic materials, by making use of meteoritic graphite, which is strongly ferromagnetic at room temperature. We examined ten samples of extraterrestrial graphite from a nodule in the Canyon Diablo meteorite. Graphite is the major phase in every sample but there are minor amounts of magnetite, kamacite, akaganéite, and other phases. By analysing the phase composition of a series of samples, we find that these iron-rich minerals can only account for about two-thirds of the observed magnetization. The remainder is somehow associated with graphite, corresponding to an average magnetization of 23 $Am^2kg^{-1}$, or 0.05 Bohr magnetons ($\mu_B$) per carbon atom. The magnetic ordering temperature is near 570 K. We suggest that the ferromagnetism is a magnetic proximity effect induced at the interface with magnetite or kamacite inclusions.**

Many carbon-based ferromagnets order magnetically below 20 K[6,7], but the recent study[4] of polymerized rhombohedral $C_{60}$ found a Curie temperature of about 500 K and a spontaneous magnetization $\sigma_S$ = 0.09 $Am^2kg^{-1}$. An extensive study of graphite samples of different provenance[1] suggested an intrinsic origin for the ferromagnetism, although the spontaneous magnetization at room temperature did not exceed 0.003 $Am^2kg^{-1}$. Another intriguing report is that ferromagnetism with $\sigma_S = 0.02$ $Am^2kg^{-1}$ may coexist with superconductivity in the graphite-sulphur system[8]. To put the weakness of the ferromagnetism in perspective, carbon with a ferromagnetic moment of 1 $\mu_B$ per atom would have $\sigma_S = 465$ $Am^2kg^{-1}$, and a corresponding polarization $J_S = 1.2$ T. It seems that only a tiny fraction of the carbon atoms participate in the magnetism of these materials, or else the carbon moment must be remarkably weak ($\sim 10^{-4}$ $\mu_B$). An exception is amorphous carbon prepared by direct pyrolysis[3], which in one case[9] was reported to have a room-temperature magnetization of 9.2 $Am^2kg^{-1}$, or 0.02 $\mu_B$ per carbon. We have reproduced this result, but find iron in the form of micron-sized oxide particles dispersed throughout the carbon.

The impact some 50,000 years ago of the 50,000-tonne Canyon Diablo meteorite at a relative velocity of about 20 km s$^{-1}$ was a cataclysmic event, which created a crater of diameter 1.3 km in the Arizona desert. Canyon Diablo is classified as a silicate-bearing IAB iron[10,11]. The origin of this group is puzzling, but it may have involved catastrophic mixing of the molten iron core of an asteroid in a collision with a chondritic body in the first few million years of the Solar System. The IAB group exhibit enormous heterogeneity in their range of inclusions. Rounded graphite-rich nodules are a particular feature of Canyon Diablo. The specimens that we have studied were taken from the nodule illustrated in Fig 1. Graphite is the major phase, but the nodule contains primary kamacite ($Fe_{94}Ni_6$) and magnetite ($Fe_3O_4$), together with traces of schreibersite ($Fe_{2.0}Ni_{1.0}$)P, troilite (FeS) and enstatite ($Mg_2SiO_3$). The metal has been locally oxidized to magnetite and akaganéite (Cl-containing $\beta$FeO(OH)), with traces of haematite ($Fe_2O_3$) in some specimens. These phases replace the metal, and also fill stress fractures induced by expansion during terrestrial weathering.

Ten graphite samples of 100 - 500 mg were investigated. Overall chemical analysis gave (in wt%) gave C 51.8; S 0.09; P 0.12; $FeO_{1.12}$ 41.6 (the average iron oxidation state comes from Mössbauer data); NiO 2.50; CoO 0.21; $SiO_2$ 1.70; $Al_2O_3$ 0.19; MnO < 0.01; MgO 0.49; CaO 0.06; $Na_2O$ 0.02; $K_2O$ 0.02; $TiO_2$ 0.01. Densities of 2400 - 3600 kg $m^{-3}$ are all greater than the X-ray density of graphite (2260 kg $m^{-3}$), which is consistent with the presence of the phases seen in Fig1. Part of each sample was ground to powder in an agate mortar for chemical and magnetic analysis. Any obvious kamacite nuggets were removed when grinding the powder. Small fragments (1-10 mg) were selected from several of the samples in order to investigate the variability of the magnetization.

Each part of every sample is strongly magnetic, jumping to a small ferrite magnet. Magnetization curves are all similar to the one illustrated in Fig 2. Values of $\sigma_S$ for the ten powder samples range from 21 - 70 $Am^2kg^{-1}$. The range of magnetization of the milligram fragments was 6 - 185 $Am^2kg^{-1}$, the latter being measured on a kamacite nugget. The magnetization data are summarized in Table I. Curie temperatures were determined from thermogravimetric scans in a magnetic field gradient (Fig 2). The two sharp transitions at 1030 K and 860 K correspond to $T_C$ for kamacite and magnetite respectively, and the broad decrease in magnetization at around 570 K is attributed to ferromagnetic graphite.

In order to estimate the magnetization associated with ferromagnetic impurity phases dispersed throughout the graphite, we characterised these phases using Mössbauer spectroscopy, together with chemical analysis and scanning electron microscopy with energy-dispersive X-ray analysis (EDAX). X-ray diffraction (Fig 3) provides a qualitative overview. Of the iron-rich minerals, magnetite, kamacite and schreibersite are intrinsically ferromagnetic or ferrimagnetic, troilite and haematite are antiferromagnetic and akaganéite is paramagnetic at room temperature. There is no doubt that the first three minerals contribute to the magnetization of meteoritic graphite, but the question is whether they are present in sufficient quantity to explain the magnetization data in Table I.

Typical Mössbauer spectra are shown in Fig 4. Besides the magnetically split components due to magnetite and kamacite and the paramagnetic akaganéite doublet, there is also a broad, poorly-resolved component with hyperfine field $B_{hf} \approx 20$ T, which could be due to poorly-crystallized ferric hydroxide or possibly cohenite ($Fe_3C$). However, low temperature spectra are fully-split, with no component with $B_{hf} > 30$ T (Fig 4b). We assume that the recoilless fractions of all phases are similar, so that the relative amounts of iron in the phases are proportional to their Mössbauer absorption areas.

In Table I, the concentration (in wt%) of the ferromagnetic minerals in each of the graphitic samples is listed, together with their combined contribution to the magnetization, $\sigma_{imp}$, calculated by assuming values of 75, 216 and 100 $Am^2kg^{-1}$ for oxidised magnetite, kamacite and schreibersite, respectively. The concentration of these minerals was derived by distributing the total iron among the phases in proportion to their Mössbauer absorption areas (or in the case of $(Fe_{2.0}Ni_{1.0})P$, based on the P content). Experimental uncertainties are given in parentheses. Assuming that the broad component is entirely due to a nanocrystalline iron phase with magnetization of 100 $Am^2kg^{-1}$, the observed magnetization, $\sigma_s$, in Table I significantly exceeds $\sigma_{imp}$ in six out of ten samples (Fig 4). The average value of $\sigma_s$ weighted by sample mass (40.6 $Am^2kg^{-1}$) is 38 % greater than that of $\sigma_{imp}$ (29.7 $Am^2kg^{-1}$). Attributing the difference $\Delta\sigma = \sigma_s - \sigma_{imp}$ to the graphite, we find an average magnetization $\sigma_c$ of 23.1 $Am^2kg^{-1}$, which corresponds to 0.05 $\mu_B$ per carbon atom. If the broad component is due to an antiferromagnetic phase, these numbers are 40 % greater.

We emphasize that our claim that the magnetization of the graphitic material significantly exceeds anything that can be attributed to the iron-rich phases does not rest on the details of the phase analysis. For example, sample 1.7 contains 44 wt. % iron according to the chemical analysis. Of this, only 38 % is magnetically ordered. Even if it was all pure iron ($\sigma_s$ = 220 Am$^2$kg$^{-1}$), the magnetization would only be 37 Am$^2$kg$^{-1}$, compared with the 52 Am$^2$kg$^{-1}$ measured on the same powder. The density, $\rho$, sets another limit of 38 Am$^2$kg$^{-1}$ for this sample, assuming a mixture of graphite and iron. Furthermore, extrapolation of $\sigma_s$ vs $\rho$ for 18 specimens gives a completely independent value of the average magnetization, 21 (13) Am$^2$kg$^{-1}$ (standard deviation in parentheses) at the density of graphite. A series of acid treatments and density separations failed to eliminate the magnetic moment.

So how does this ferromagnetism come about? One idea is that meteoritic graphite is somehow different from terrestrial graphite, due perhaps to its mode of formation, chemical doping (Fe, S, P, N…), or the shock of impact. Defects tend to enhance the susceptibility of graphite, and may even lead to superconductivity[12]. There are theoretical reports that graphene strips a few nanometers in width show an unusual density of states which leads to local moments at the strip edges which are paramagnetic[13], or, with a suitable stacking, antiferromagnetic[14]. But there have been no suggestions that nanographite should be ferromagnetic. Ferromagnetism would not normally be expected in graphite, a semi-metal with Fermi energy $E_F \approx 20$ meV, where the Fermi level falls at a sharp minimum in the density of states[15]. Nor is there any indication on our X-ray patterns of intercalation or poor crystallinity on the nanometer scale. The unit cell is about 1 % shorter in both $a$ and $c$ directions than usual, with lattice parameters $a_0$ = 0.2450(5) nm, $c$ = 0.6713(5) nm, but neither c-axis nor basal-plane reflections are anomalously broad. EDAX analysis of the graphite regions shows no heavy element other than iron at the 0.1 wt% level.

Another idea is that the ferromagnetic inclusions somehow induce a moment in the graphite. Scanning electron micrographs indicated that magnetite and kamacite particles of all sizes down to 50 nm or less are individually embedded in graphite (Fig 1). Perfectly-encapsulated particles are not removed by acid treatment. Since magnetite is a half-metal[16], matching the chemical potential of the ↑ and ↓ electrons at the interface leads to complete spin polarization of the adjacent carbon atoms. If the spin splitting of the graphite bands at the $Fe_3O_4$ interface is a few tenths of an electron volt, the Fermi level there falls in a higher density of states, tending to sustain the spin polarization. The recently-discovered gapless spin density mode in graphene sheets[5] may greatly enhance the spin susceptibility. Accounting for the magnetization of the graphite in a model where spherical inclusions of radius $r_o$ are dressed by a spin polarization decaying as $\exp\{-(r-r_o)/\lambda_s\}$, the ratio of the moment of the inclusion and the magnetic carbon shell is approximately $1/3\rho(1 + 2\rho + 2\rho^2)$ where $\rho = \lambda_s/r_o$, and $\lambda_s$ is the spin decay length. Taking $\sigma_{imp}/\Delta\sigma = 2.8$, a typical value of $r_o = 50$ nm gives $\lambda_s = 5$ nm. The Fermi wavelength for graphite is 10.6 nm.

The magnetoresistance of the meteoritic graphite shows a positive quadratic variation with applied B field, reaching 2.9 % in a field of 2 T at room temperature and 4.8 % at 5 K. The in-plane mobility $\mu$ in a two-band model with equal concentrations $n$ of electrons and holes is $(\Delta\rho/\rho B^2)^{1/2} \approx 0.1$ m$^2$ V$^{-1}$s$^{-1}$. Taking $n = 7$ x $10^{24}$ m$^{-3}$ [15], the mean free path $\lambda = (\hbar\mu/e)(3\pi^2 n)^{1/3}$ is estimated as 40 nm, and the spin-diffusion length, $\lambda_{sd}$, is expected to be an order of magnitude greater. In carbon nanotubes, $\lambda_{sd}$ is known to exceed 100 nm[17]. Furthermore, there is no resistivity mismatch for a cubic particle embedded in graphite provided its resistivity lies between the graphite c-axis and a-axis values, as is the case for magnetite. Provided that spin injection can be achieved, the combination of half-metal (magnetite) and semi-metal (graphite) looks promising for spin electronic applications.

Ferromagnetic carbon produced by pyrolysis of organic precursors should be reinvestigated using the present methods, in order to assess if it too is ferromagnetic because of a magnetic proximity effect, as we propose here for meteoritic graphite. The implications of ferromagnetic carbon, whatever its origin, are likely to be wide-reaching – this material could be a zero-gap, high-temperature ferromagnetic semiconductor. With this new sighting of strongly ferromagnetic meteoritic carbon, we look forward to rapid developments in the area of carbon-based magnetism..

1. Esquinazi P., Setzer A., Höhne R., Semmelhack C., Kopelevich Y., Spemann D., Butz T., Kohlstrunk B. & Loesche M. Ferromagnetism in oriented graphite samples. Phys. Rev. B 66, 0244429 (2002).

2. Höhne R. & Esquinazi P. Can carbon be ferromagnetic? *Adv. Mater.* **14**, 753 - 756 (2002).

3. Makarova T. L. Magnetism of carbon-based materials, *Studies of High-Tc Superconductivity* (ed. Narlikar, T.) **44-45** (Nova Science Publishers)

4. Makarova T. L., Sundqvist B., Höhne R., Esquinazi P., Kopelevich Y., Scharff P., avydov V. A., Kashevarova L. S. & Rakhmanina A. V., Magnetic carbon. *Nature* **413**, 716 - 718 (2001).

5. Baskaran G. & Jafari S. A., Predicting a Gapless Spin-1 Neutral Collective Mode branch for Graphite. *Phys. Rev. Lett.* **89** 016402 (2002).

6. Allemand P. M., Khemani K. C., Koch A., Wudl F., Holczer K., Donovan S., Gruner G. & Thompson J. D. Organic molecular soft ferromagnetism in a fullerene $C_{60}$. *Science* **253**, 301 - 303 (1991).

7. Mrzel A., Omerzu A., Umek P., Mihailovic D., Jagličić Z. & Trontelj Z. Ferromagnetism in a cobaltocene-doped fullerene derivative below 19 K due to unpaired spins on fullerene molecules. *Chem. Phys. Lett.* **298**, 329 -334 (1998).

8. Moehlecke S., Ho P. C. & Maple M. B., Coexistence of superconductivity and ferromagnetism in the graphite-sulfur system. *Phil. Mag. B* **82**, 1335-1347 (2002)

9. Murata K., Ushijima H., Ueda H. & Kawaguchi K., A stable carbon-based organic magnet. *J. Chem Soc., Chem. Commun.* **7**, 567 - 569 (1992).

10. Grady M. M., *Catalogue of Meteorite* 5th edition (Natural History Museum, London) pp 89 - 90 (2000).

11. Papike J. J., (editor) *Planetary Materials* Reviews in Mineralogy vol 35 (The mineralogical Society of America, Washington DC) 1998.

12. González J., Guinea F. & Vozmediano M. A. H., Electron-electron interactions in graphene sheets. *Phys. Rev. B* **63** 134421 (2001).

13. Wakabayashi K., Jujita M., Ajiki H. & Sigrist M, Magnetic properties of nanographites at low temperature. *Physica B* **280**, 388 - 389 (2000).

14. Harigaya K., The mechanism of magnetism in stacked nanographite: theoretical study. *J. Phys.: Condens. Matter* **13**, 1295 - 1302 (2001).

15. Kelly B. T., Physics of Graphite. *Applied Science Publishers*, London (1981).

16. Coey J. M. D. & Venkatesan M., Half-metallic ferromagnets; the example of $CrO_2$. *J. Appl. Phys* **91**, 8345 - 8350 (2002).

17. Alphenaar B. W., Tsukagoshi K & Ago H, Spin electronics using carbon nanotubes. *Physica* E **6**, 848 - 851 (2000).

18. Brett R. & Higgins G. T., Cliftonite: A proposed origin and its bearing on the origin of diamonds in meteorites. *Geochim. Cosmochim. Acta* **33,** 1473-1484 (1969).

Acknowledgements. We thank M. Viret, F. Guinea and  S. Sanvito for discussions. This work was supported by Science Foundation Ireland.

Correspondence and requests for materials should be addressed to J.M.D.C. (e-mail: jcoey@tcd.ie).

Figure captions.

Fig 1.  The Canyon Diablo graphite nodule at increasing magnifications: (a) Cut and polished section of the whole nodule showing interconnecting veins of kamacite ($Fe_{94}Ni_6$) in the graphite matrix; (b) Backscattered electron image of graphite lightly peppered with tiny nuggets of kamacite (bright flecks) and traversed by thin, subparallel veins of oxidized iron (pale grey). A partially oxidised kamacite vein (bottom right) contains black inclusions of cliftonite[18], a variety of graphite comprised of radiating clusters of crystallites that appear to have grown by exsolution from carbon-rich metal on cooling;    (c) Reflected polarised light image of polished graphite showing three distinct forms: A, a continuous border of cliftonite surrounding an embayed nugget of kamacite, B, a  large buckled plate of graphite (pale gold), and C, a poorly polished, felted mass of graphite crystallites.  Inset, a cliftonite inclusion in metal at the same scale; (d) Backscattered image of graphite with embedded iron-rich inclusions, possibly magnetite, down to 50 nm in size.

Figure 2. Typical room-temperature curve of magnetization $\sigma$ against applied field $\mu_0 H$ for graphitic material from the Canyon Diablo meteorite. Insets display the hysteresis (top left) and a thermomagnetic scan showing the Curie temperatures of graphite, magnetite and kamacite (bottom right). There is a little hysteresis, with coercivity $\mu_0 H_C$ in the range 6 - 8  mT. The low-field tangent to the magnetization curves intersects the line $\sigma = \sigma_S$ at a field $\mu_0 H \approx 200$ mT, which is consistent with the presence of spherical particles with a demagnetizing factor $N \approx 1/3$ and $J_s \approx 0.6$ T.

Figure 3) Typical X-ray diffraction pattern for graphitic material from the Canyon Diablo meteorite. The peaks are assigned to graphite/cliftonite (C), magnetite (M), kamacite (K), akaganéite (A) and haematite (H).

Figure 4) Ferromagnetic phase analysis of samples of graphitic material from the Canyon Diablo meteorite. **a, b** Typical Mössbauer spectra at room temperature (a) and at 15 K (b). The contributions from the A and B sites of magnetite, kamacite, enstatite and akaganéite are denoted by $M_A$, $M_B$, K, E and A, respectively. From the intensity ratio of $M_A$ and $M_B$, the magnetic sextets with hyperfine fields $B_{hf}$ of 49.0(5) T and 45.8(5) T, we deduce that the magnetite is oxidized, with composition $Fe_{2.83}O_4$. The lattice parameter is $a_0 = 0.8351(5)$ nm. The central paramagnetic doublet with isomer shift $\delta = 0.36(1)$ mm s$^{-1}$ relative to $\alpha$-Fe at 300 K and quadrupole splitting $\Delta = 0.69(2)$ mm s$^{-1}$ is associated with akaganéite. This component is not superparamagnetic in an applied field of 110 mT, as there is no change in the intensity or linewidth of the doublet. The weaker, ferrous doublet (E) with $\delta = 1.17(1)$ mm s$^{-1}$ and $\Delta = 2.00(2)$ mm s$^{-1}$ is due to iron in M2 sites of enstatite. In most samples, there is the component (K) with $B_{hf} = 33.5(5)$ T and $\delta = 0.04(1)$ mm s$^{-1}$ which is associated with kamacite. A component with $B_{hf} = 31.0(5)$ T observed in one case is associated with troilite. **c** Contribution to the magnetisation of the ferromagnetic phases (shaded: nano-crystalline iron (nc iron) is probably overestimated, as this phase is likely to be antiferromagnetic). The unshaded residue is associated with graphite.

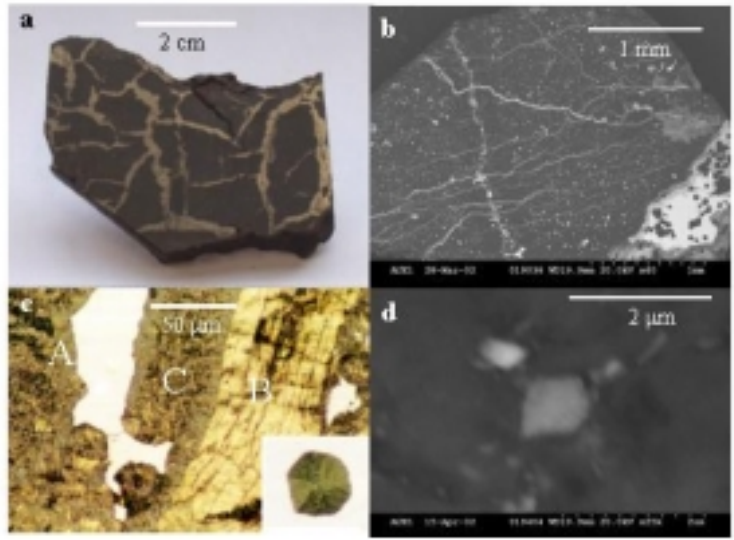

Figure 1

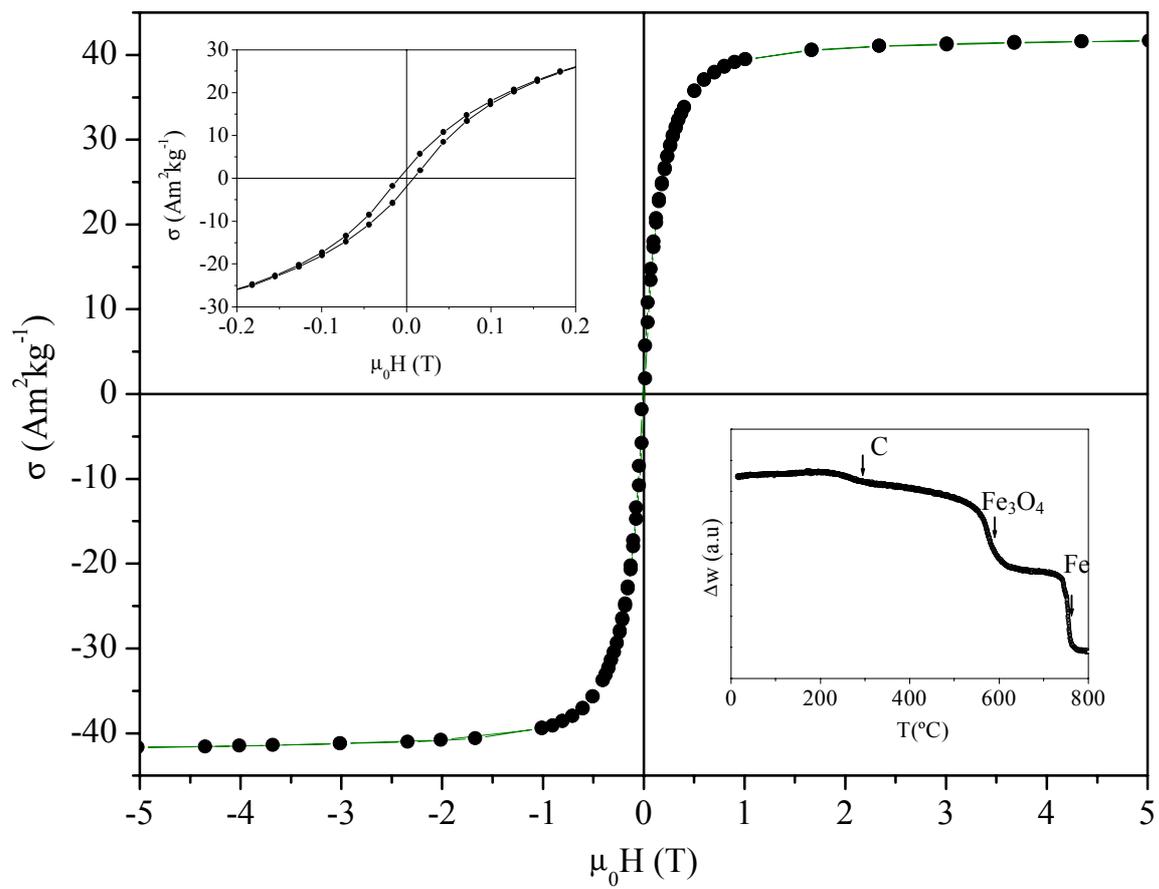

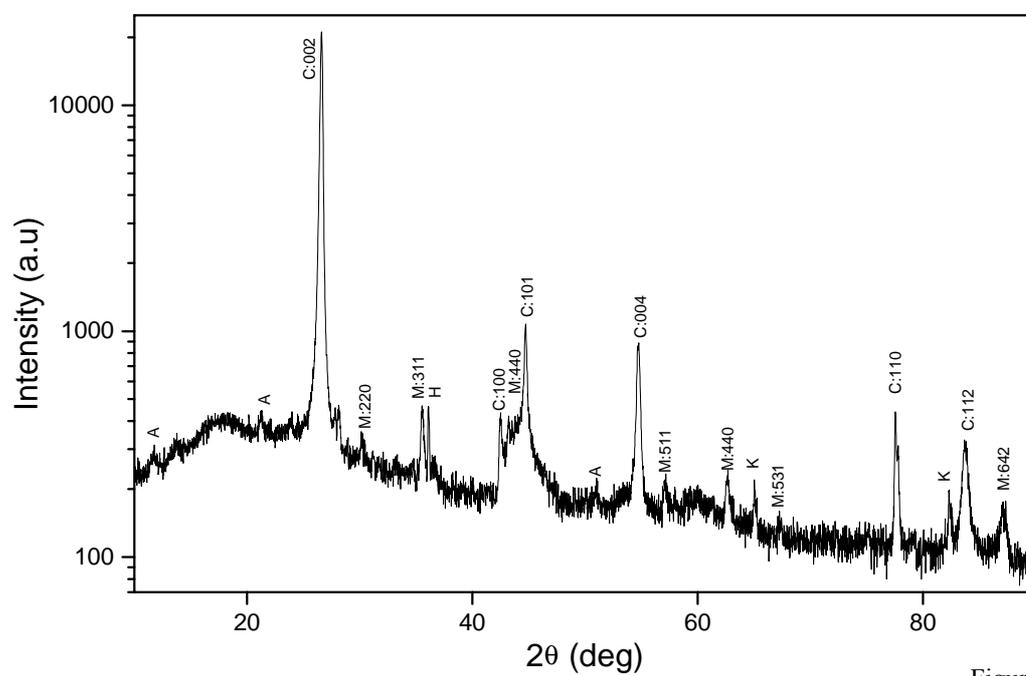

Figure 3

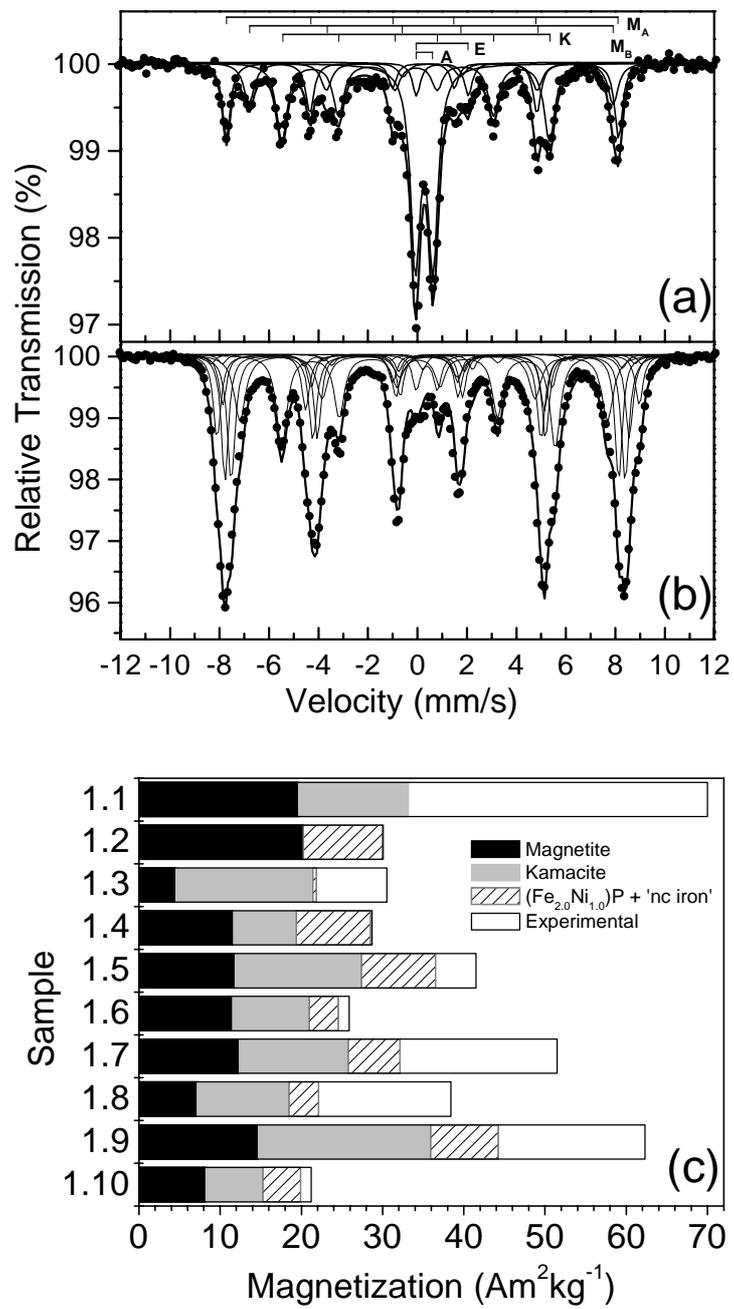

Figure 4

Table I **Ferromagnetic properties and related data for samples from the Canyon Diablo graphite**

| Sample | $\rho$ (kg m$^{-3}$) | $\sigma_s$ (Am$^2$kg$^{-1}$) | LOI % | C (wt.%) | Fe (wt.%) | Fe$_3$O$_4$ (wt.%) | Fe$_{94}$Ni$_{16}$ (wt.%) | (Fe$_{2.0}$Ni$_{1.0}$)P 'nc iron' (wt.%) | $\sigma_{imp.}$ (Am$^2$kg$^{-1}$) | $\Delta\sigma$ (Am$^2$kg$^{-1}$) | $\sigma_c$ (Am$^2$kg$^{-1}$) |
|---|---|---|---|---|---|---|---|---|---|---|---|
| 1.1 | nd | 70.0 | 51.8 | 54.2 | 31.8 | 26.2 | 6.3 | 0.0 | 33.3 | 36.7 | 67.7 |
| 1.2 | nd | 30.0 | 43.9 | 48.5 | 36.4 | 27.0 | 0.0 | 9.8 | 30.1 | - | -0.2 |
| 1.3 | 2630 | 30.5 | 74.9 | 69.8 | 18.8 | 6.1 | 7.8 | 0.4 | 21.8 | 8.7 | 12.5 |
| 1.4 | 3040 | 28.7 | 46.5 | 51.5 | 32.0 | 15.6 | 3.6 | 9.1 | 28.6 | 0.1 | 0.2 |
| 1.5 | 3010 | 41.5 | 46.2 | 46.2 | 43.5 | 15.8 | 7.2 | 9.1 | 36.6 | 4.9 | 10.6 |
| 1.6 | 2430 | 25.9 | 62.6 | 66.1 | 23.3 | 15.4 | 4.3 | 3.6 | 24.5 | 1.4 | 2.1 |
| 1.7 | 3350 | 51.5 | 28.5 | 39.8 | 44.0 | 16.5 | 6.2 | 6.4 | 32.2 | 19.3 | 48.5 |
| 1.8 | 3380 | 38.4 | 51.8 | 54.4 | 29.0 | 9.6 | 5.2 | 3.6 | 22.0 | 16.4 | 30.1 |
| 1.9 | 3360 | 62.3 | 29.0 | 36.7 | 44.8 | 19.6 | 9.8 | 8.3 | 44.2 | 18.1 | 49.3 |
| 1.10 | 3650 | 21.2 | 53.8 | 56.0 | 28.5 | 11.0 | 3.2 | 4.6 | 19.8 | 1.4 | 2.5 |
| **Average (SD)** | **3106 (412)** | **40.6 (13.1)** | | **52.1 (11.7)** | **34.2 (10.5)** | | | | **29.7 (7.0)** | **10.8 (9.5)** | **23.1 (21.5)** |

The averages are weighted by sample mass. 'nc iron', nanocrystalline iron; $\rho$, density; $\sigma_s$, spontaneous magnetisation; LOI, loss on ignition; $\sigma_{imp.}$, total magnetisation of ferromagnetic inclusions; $\Delta\sigma = \sigma_s - \sigma_{imp.}$; $\sigma_c$, magnetisation attributed to carbon. ND, not determined.